\documentclass[10pt,reqno]{amsart}
     \makeatletter
     \def\section{\@startsection{section}{1}%
     \z@{.7\linespacing\@plus\linespacing}{.5\linespacing}%
     {\bfseries
     \centering
     }}
     \def\@secnumfont{\bfseries}
     \makeatother
\usepackage{amsmath}
\usepackage{amssymb}
\usepackage[latin1]{inputenc}
\usepackage{bbm}
\newcommand*{\cA}{\mathcal{A}}
\newcommand*{\cB}{\mathcal{B}}
\newcommand*{\cC}{\mathcal{C}}
\newcommand*{\C}{\mathbb{C}}

\newcommand*{\N}{\mathbb{N}}
\newcommand*{\1}{\mathbbm{1}}
\newcommand*{\R}{\mathbb{R}}

\newcommand{\abs}[1]{\left|#1\right|}
\newcommand{\norm}[1]{\left\|#1\right\|}

\renewcommand{\and}{\text{ and }}

\newtheorem{theorem}{Theorem}[section]
\newtheorem{lemma}[theorem]{Lemma}

\newtheorem{corollary}[theorem]{Corollary}
\theoremstyle{definition}
\newtheorem{definition}[theorem]{Definition}
\newtheorem{example}[theorem]{Example}
\theoremstyle{remark}
\newtheorem{remark}[theorem]{Remark}
\numberwithin{equation}{section}
\begin{document}

\title[Momentum Space Path Integrands]{Hamiltonian Path Integrals in Momentum Space Representation via White Noise Techniques}

\author[Wolfgang Bock]{Wolfgang Bock}
\address{Functional Analysis and Stochastic Analysis Group, \\
Department of Mathematics, \\
University of Kaiserslautern, 67653 Kaiserslautern, Germany}
\email{bock@mathematik.uni-kl.de}
\urladdr{http://www.mathematik.uni-kl.de/$\sim$bock}



\begin{abstract}
The concepts of Feynman integrals in white noise analysis are used to construct the Feynman integrand for the harmonic oscillator in momentum space representation as a Hida distribution. Moreover it is shown that in a limit sense, the potential free case fulfills the conservation of momentum. 
\end{abstract}

\maketitle

\section{Introduction}
The Feynman path integral is a very successfully applied object. Although the first aim of Feynman was to develop path integrals based on a Lagrangian, they also can be used for various systems which have a law of least action, see e.g.\cite{F48}.\\
Since classical quantum mechanics is based on a Hamiltonian formulation rather than a Lagrangian one, it is worthwhile to take a closer look to the so-called Hamiltonian path integral, which means the Feynman integral in phase space.
Feynman gave a heuristic formulation of the phase space Feynman Integral
\begin{equation}\label{psfey}
K(t,y|0,y_0)= {\rm N} \int_{x(0)=y_0, x(t)=y} \int \exp\left(\frac{i}{\hbar} S(x)\right) \prod_{0<\tau<t}  \frac{dp(\tau)}{(2\pi)^d} dx(\tau)
\end{equation}
in \cite{Fe51}.
Here the action (and hence the dynamic) is expressed by a canonical (Hamiltonian) system of generalized space variables $x$ and their corresponding conjugate momenta $p$. The canonical variables can be found by a Legendre-transformation, see e.g.~\cite{Sch07}.
The Hamiltonian action:
\begin{equation*}
S(x,p,t)=\int_0^t p(\tau)\dot{x}(\tau) -H(x(\tau),p(\tau),\tau) d\tau,
\end{equation*}
where 
\begin{equation*}
H(x,p,t)=\frac{1}{2m}p^2 +V(x,p,t)
\end{equation*}
is the Hamilton function and given by the sum of the kinetic energy and the potential.
Note that both integrals, the Feynman integral as well as the Hamiltonian path integral are thought to be integrals w.r.t~a flat, i.e.~translation invariant measure on the infinite dimensional path space. Such a measure does not exist, hence the integral in \ref{psfey} is not a mathematical rigorous object. Nevertheless there is no doubt that it has a physical meaning.\\
The Hamiltonian setting has many advantages such as e.g.~:
\begin{itemize}
\item the semi-classical limit of quantum mechanics is more natural in an Hamiltonian setting, i.e. the phase space is more natural in classical mechanics than the configuration space, see also \cite{AHKM08, KD82} and the references therein.
\item in \cite{DK83} the authors state, that potentials which are time-dependent or velocity dependent should be treated with the Hamiltonian path integral. 
\item momentum space propagators can be investigated.
\end{itemize}
There are many attempts to give a meaning to the Hamiltonian path integral as a mathematical rigorous object. Among these are analytic continuation of probabilistic integrals via coherent states \cite{KD82, KD84} and infinite dimensional distributions e.g.~\cite{DMN77}. Another approach by Albeverio et al.~uses Fresnel integrals e.g.~\cite{AHKM08, AGM02} and most recently a method using time-slicing was developed by Naoto Kumano-Go \cite{Ku11}. 
As a guide to the literature on many attempts to formulate these ideas we point out the list in \cite{AHKM08}.\\
In this article we choose a White Noise approach to give a meaning to Hamiltonian integrands in momentum space representation as distributions of White Noise Analysis. \\
White Noise Analysis is a mathematical framework which offers generalizations of concepts from finite-dimensional analysis, like differential operators and Fourier transform to an infinite-dimensional setting. We give a brief introduction to White Noise Analysis in Section 2, for more details see \cite{Hi80,BK95,HKPS93,Ob94,Kuo96}. Of special importance in White Noise Analysis are spaces of generalized functions and their characterizations. In this article we choose the space of Hida distributions, see Section 2.\\
The idea of realizing Feynman integrals within the White Noise framework goes back to \cite{HS83}. As ansatz for the Feynman integrand in configuration space one has
\begin{multline}\label{integrandpot}
I_V = {\rm Nexp}\left( \frac{i}{2\hbar}\int_0^t \dot{x}^2(\tau)d\tau +\frac{1}{2}\int_0^t \dot{x}(\tau)^2 d\tau\right)\\
\times \exp\left(-\frac{i}{\hbar} \int_0^t V(x(\tau),\dot{x}(\tau)) \, d\tau\right) \cdot \delta(x(t)-y),
\end{multline}
where $x(t)=y_0+ B_t$ is a Brownian path starting in $y_0$. In equation \eqref{integrandpot} the first integral in the exponential represents the kinetic energy and the second integral the compensation of the Gaussian fall-off. The delta function (Donsker's Delta) pins the paths at the end time point in the end point. The normalized exponential as well as the delta function exists as well-defined objects in the space of Hida distributions.
With this concept many authors constructed the Feynman integrand for a large class of Lagrange functions and thus the Feynman integral as generalized expectation of the integrand w.r.t.~ the Gaussian measure see e.~g.~\cite{BCB02}, \cite{FPS91}, \cite{FOS05},
\cite{GKSS97}, \cite{HKPS93}, \cite{KS92}, \cite{KWS97},
\cite{L06}, \cite{SS04}, \cite{BGJ13}.\\
In \cite{BG11} the concepts from \cite{GS98a} are used to give a mathematical rigorous meaning to the Hamiltonian path integrand as a Hida distribution in the case of non-velocity dependent potentials.\\
In momentum space representation we know about the initial and the end momenta, it is clear by Heisenberg uncertainty principle that we have no certain information about the corresponding space variables. This means we model the momentum trajectories as a Brownian fluctuation starting in the initial momentum $p_0$. 
\begin{equation}\label{mommom}
p(\tau)=p_0+\frac{\sqrt{\hbar m}}{t-t_0}B(\tau),\quad 0\leq \tau \leq t.
\end{equation}
Furthermore the space variable is modeled by white noise, i.e.
\begin{equation}\label{mompath}
x(\tau) =\sqrt{\frac{\hbar}{m}}\cdot(t-t_0)\omega_x(\tau),\quad 0\leq \tau \leq t. 
\end{equation}
This is a meaningful definition, since a path has always start and end points which a noise does not have. Moreover since we have that if the initial and end conditions of the momenta are fully known, the space variable is completely uncertain, which means has variance infinity. The white noise process is intrinsically fulfilling the no boundary condition property and has as well infinite variance.

The Hamiltonian path integral for the momentum space propagator is formally given by, see e.g.~\cite{K04}
\begin{equation}\label{Hampathmom}
K(p',t',p_0,t_0)=\mathcal{N} \int_{p({t_0})=p_0, p(t)=p'} \exp(\frac{i}{\hbar} \int_{t_0}^{t} -q(s) \dot{p}(s)-H(p,q) \ ds) \,Dp Dq.
\end{equation}
This path integral can be obtained by a Fourier transform of the coordinate space path integral in both variables, see e.g.~\cite{Kl03}.
Then we propose the following formal ansatz for the Feynman integrand in Phase space with respect to the Gaussian measure $\mu$,
\begin{align}\label{anpsfeymom}
I_V = {\rm N}\exp\left( \frac{i}{\hbar}\int_{t_0}^t  -x(\tau) \dot{p}(\tau) -\frac{p(\tau)^2}{2m} d\tau +\frac{1}{2}\int_{t_0}^t \omega_x(\tau)^2 +\omega_p(\tau)^2 d\tau\right)\\ \nonumber
\times \exp\left(-\frac{i}{h} \int_{t_0}^t V(x(\tau),p(\tau),\tau) \, d\tau\right) \cdot \delta(p(t)-p').
\end{align}
In this expression the sum of the first and the third integral is the action $S(x,p)$ in momentum representation, and the Donsker's delta function serves to pin trajectories to $p'$ at time $t$. The second integral is introduced to simulate the Lebesgue integral by compensation of the fall-off of the Gaussian measure in the time interval $({t_0},t)$. Furthermore, as in Feynman's formula we need a normalization which turns out to be infinity and will be implemented by the use of a normalized exponential as in Chapter \ref{GGK}.
We use the concepts of quadratic actions in white noise analysis, which were further developed in \cite{GS98a} and \cite{BG11} to give a rigorous meaning to the Feynman integrand $I_V$ in \ref{anpsfeymom} as a White Noise distribution.
The construction is done in terms of the $T$-transform (infinite-dimensional version of the Fourier transform w.r.t~a Gaussian measure), which characterizes Hida distributions, see Theorem \ref{charthm}. At the same time, the $T$-transform of the constructed Feynman integrands provides us with their generating functional. Finally using the generating functional, we can show that the generalized expectation (generating functional at zero) gives the Green's function to the corresponding Schrödinger equation. 

Note that both integrals, the Feynman integral as well as the Hamiltonian path integral are thought to be integrals w.r.t~a flat, i.e.~translation invariant measure on the infinite dimensional path space. Such a measure does not exist, hence the integral at first - as it stands - is not a mathematical rigorous object. The normalization constant in both integrals turns out to be infinity. Nevertheless there is no doubt that it has a physical meaning.\\

\section{White Noise Analysis}
\subsection{Gel'fand Triples}
Starting point is the Gel'fand triple $S_d(\R) \subset L^2_d(\R) \subset S'_d(\R)$ of the $\R^d$-valued, $d \in \N$, Schwartz test functions and tempered distributions with the Hilbert space of (equivalence classes of) $\R^d$-valued square integrable functions w.r.t.~the Lebesgue measure as central space (equipped with its canonical inner product $(\cdot, \cdot)$ and norm $\|\cdot\|$), see e.g.~ \cite[Exam.~11]{W95}.
Since $S_d(\R)$ is a nuclear space, represented as projective limit of a decreasing chain of Hilbert spaces $(H_p)_{p\in \N}$, see e.g.~\cite[Chap.~2]{RS75a} and \cite{GV68}, i.e.~
\begin{equation*}
S_d(\R) = \bigcap_{p \in \N} H_p,
\end{equation*}
we have that $S_d(\R)$ is a countably Hilbert space in the sense of Gel'fand and Vilenkin \cite{GV68}. We denote the inner product and the corresponding norm on $H_p$ by $(\cdot,\cdot)_p$ and $\|\cdot\|_p$, respectively, with the convention $H_0 = L^2_d(\R)$.
Let $H_{-p}$ be the dual space of $H_p$ and let $\langle \cdot , \cdot \rangle$ denote the dual pairing on $H_{p} \times H_{-p}$. $H_{p}$ is continuously embedded into $L^2_d(\R)$. By identifying $L_d^2(\R)$ with its dual $L_d^2(\R)'$, via the Riesz isomorphism, we obtain the chain $H_p \subset L_d^2(\R) \subset H_{-p}$.
Note that $\displaystyle S'_d(\R)= \bigcup_{p\in \N} H_{-p}$, i.e.~$S'_d(\R)$ is the inductive limit of the increasing chain of Hilbert spaces $(H_{-p})_{p\in \N}$, see  e.g.~\cite{GV68}.
We denote the dual pairing of $S_d(\R)$ and $S'_d(\R)$ also by $\langle \cdot , \cdot \rangle$. Note that its restriction on $S_d(\R) \times L_d^2(\R)$ is given by $(\cdot, \cdot )$.
We also use the complexifications of these spaces denoted with the sub-index $\C$ (as well as their inner products and norms). The dual pairing we extend in a bilinear way. Hence we have the relation 
\begin{equation*}
\langle g,f \rangle = (\mathbf{g},\overline{\mathbf{f}}), \quad \mathbf{f},\mathbf{g} \in L_d^2(\R)_{\C},
\end{equation*}
where the overline denotes the complex conjugation.
\subsection{White Noise Spaces}
We consider on $S_d' (\R)$ the $\sigma$-algebra $\cC_{\sigma}(S_d' (\R))$ generated by the cylinder sets $\{ \omega \in S_d' (\R) | \langle \xi_1, \omega \rangle \in F_1, \dots ,\langle \xi_n, \omega \rangle \in F_n\} $, $\xi_i \in S_d(\R)$, $ F_i \in \cB(\R),\, 1\leq i \leq n,\, n\in \N$, where $\cB(\R)$ denotes the Borel $\sigma$-algebra on $\R$.\\
\noindent The canonical Gaussian measure $\mu$ on $C_{\sigma}(S_d'(\R))$ is given via its characteristic function
\begin{eqnarray*}
\int_{S_d' (\R)} \exp(i \langle {\bf f}, \boldsymbol{\omega} \rangle ) d\mu(\boldsymbol{\omega}) = \exp(- \tfrac{1}{2} \| {\bf f}\|^2 ), \;\;\; {\bf f} \in S_d(\R),
\end{eqnarray*}
\noindent by the theorem of Bochner and Minlos, see e.g.~\cite{Mi63}, \cite[Chap.~2 Theo.~1.~11]{BK95}. The space $(S_d'(\R),\cC_{\sigma}(S_d'(\R)), \mu)$ is the ba\-sic probability space in our setup.
The cen\-tral Gaussian spa\-ces in our frame\-work are the Hil\-bert spaces $(L^2):= L^2(S_d'(\R),$ $\cC_{\sigma}(S_d' (\R)),\mu)$ of complex-valued square in\-te\-grable func\-tions w.r.t.~the Gaussian measure $\mu$.\\
Within this formalism a representation of a d-dimensional Brownian motion is given by 
\begin{equation}\label{BrownianMotion}
{\bf B}_t ({\boldsymbol \omega}) :=(B_t(\omega_1), \dots, B_t(\omega_d)):= ( \langle  \1_{[0,t)},\omega_1 \rangle, \dots  \langle  \1_{[0,t)},\omega_d \rangle),\end{equation}
with ${\boldsymbol \omega}=(\omega_1,\dots, \omega_d) \in S'_d(\R),\quad t \geq 0,$
in the sense of an $(L^2)$-limit. Here $\1_A$ denotes the indicator function of a set $A$. 

\subsection{The Hida triple}

Let us now consider the Hilbert space $(L^2)$ and the corresponding Gel'fand triple
\begin{equation*}
(S) \subset (L^2) \subset (S)'.
\end{equation*}
Here $(S)$ denotes the space of Hida test functions and $(S)'$ the space of Hida distributions. In the following we denote the dual pairing between elements of $(S)$ and $(S)'$ by $\langle \! \langle \cdot , \cdot \rangle \!\rangle$. 
Instead of reproducing the construction of $(S)'$ here we give its characterization in terms of the $T$-transform.\\
\begin{definition}
We define the $T$-transform of $\Phi \in (S)'$ by
\begin{equation*}
T\Phi({\bf f}) := \langle\!\langle  \exp(i \langle {\bf f}, \cdot \rangle),\Phi \rangle\!\rangle, \quad  {\bf f}:= ({ f_1}, \dots ,{ f_d }) \in S_{d}(\R).
\end{equation*}
\end{definition}

\begin{remark}
\begin{itemize}
\item[(i)] Since $\exp(i \langle {\bf f},\cdot \rangle) \in (S)$ for all ${\bf f} \in S_d(\R)$, the $T$-transform of a Hida distribution is well-defined.
\item[(ii)] For ${\bf f} = 0$ the above expression yields $\langle\!\langle \Phi, 1 \rangle\!\rangle$, therefore $T\Phi(0)$ is called the generalized expectation of $\Phi \in (S)'$.
\end{itemize}
\end{remark}

\noindent In order to characterize the space $(S)'$ by the $T$-transform we need the following definition.

\begin{definition}
A mapping $F:S_{d}(\R) \to \C$ is called a {\emph U-functional} if it satisfies the following conditions:
\begin{itemize}
\item[U1.] For all ${\bf{f, g}} \in S_{d}(\R)$ the mapping $\R \ni \lambda \mapsto F(\lambda {\bf f} +{\bf g} ) \in \C$ has an analytic continuation to $\lambda \in \C$ ({\bf{ray analyticity}}).
\item[U2.] There exist constants $0<C,D<\infty$ and a $p \in \N_0$ such that 
\begin{equation*}
|F(z{\bf f})|\leq C\exp(D|z|^2 \|{\bf f} \|_p^2), 
\end{equation*}
for all $z \in \C$ and ${\bf f} \in S_{d}(\R)$ ({\bf{growth condition}}).
\end{itemize}
\end{definition}

\noindent This is the basis of the following characterization theorem. For the proof we refer to \cite{PS91,Kon80,HKPS93,KLPSW96}.

\begin{theorem}\label{charthm}
A mapping $F:S_{d}(\R) \to \C$ is the $T$-transform of an element in $(S)'$ if and only if it is a U-functional.
\end{theorem}
Theorem \ref{charthm} enables us to discuss convergence of sequences of Hida distributions by considering the corresponding $T$-transforms, i.e.~ by considering convergence on the level of U-functionals. The following corollary is proved in \cite{PS91,HKPS93,KLPSW96}.

\begin{corollary}\label{seqcor}
Let $(\Phi_n)_{n\in \N}$ denote a sequence in $(S)'$ such that:
\begin{itemize}
\item[(i)] For all ${\bf f} \in S_{d}(\R)$, $((T\Phi_n)({\bf f}))_{n\in \N}$ is a Cauchy sequence in $\C$.
\item[(ii)] There exist constants $0<C,D<\infty$ such that for some $p \in \N_0$ one has 
\begin{equation*}
|(T\Phi_n)(z{\bf f })|\leq C\exp(D|z|^2\|{\bf f}\|_p^2)
\end{equation*}
for all ${\bf f} \in S_{d}(\R),\, z \in \C$, $n \in \N$.
\end{itemize}
Then $(\Phi_n)_{n\in \N}$ converges strongly in $(S)'$ to a unique Hida distribution.
\end{corollary}

\begin{example}[Vector valued white noise]
\noindent Let $\,{\bf{B}}(t)$, $t\geq 0$, be the $d$-di\-men\-sional Brow\-nian motion as in \eqref{BrownianMotion}. 
Consider $$\frac{{\bf{B}}(t+h,\boldsymbol{\omega}) - {\bf{B}}(t,\boldsymbol{\omega})}{h} = (\langle \frac{\1_{[t,t+h)}}{h} , \omega_1 \rangle , \dots (\langle \frac{\1_{[t,t+h)}}{h} , \omega_d \rangle),\quad h>0.$$ 
Then in the sense of Corollary \ref{seqcor} it exists
\begin{eqnarray*}
\langle {\boldsymbol\delta_t}, {\boldsymbol \omega} \rangle := (\langle \delta_t,\omega_1 \rangle, \dots ,\langle \delta_t,\omega_d \rangle):= \lim_{h\searrow 0} \frac{{\bf{B}}(t+h,\boldsymbol{\omega}) - {\bf{B}}(t,\boldsymbol{\omega})}{h}.
\end{eqnarray*}
Of course for the left derivative we get the same limit. Hence it is natural to call the generalized process $\langle {\boldsymbol\delta_t}, {\boldsymbol \omega} \rangle$, $t\geq0$ in $(S)'$ vector valued white noise. One also uses the notation ${\boldsymbol \omega}(t) =\langle{\boldsymbol\delta_t}, {\boldsymbol \omega} \rangle$, $t\geq 0$. 
\end{example}

Another useful corollary of Theorem \ref{charthm} concerns integration of a family of generalized functions, see \cite{PS91,HKPS93,KLPSW96}.

\begin{corollary}\label{intcor}
Let $(\Lambda, \cA, \nu)$ be a measure space and $\Lambda \ni\lambda \mapsto \Phi(\lambda) \in (S)'$ a mapping. We assume that its $T$--transform $T \Phi$ satisfies the following conditions:
\begin{enumerate}
\item[(i)] The mapping $\Lambda \ni \lambda \mapsto T(\Phi(\lambda))({\bf f})\in \C$ is measurable for all ${\bf f} \in S_d(\R)$.
\item[(ii)] There exists a $p \in \N_0$ and functions $D \in L^{\infty}(\Lambda, \nu)$ and $C \in L^1(\Lambda, \nu)$ such that 
\begin{equation*}
   \abs{T(\Phi(\lambda))(z{\bf f})} \leq C(\lambda)\exp(D(\lambda) \abs{z}^2 \norm{{\bf f}}^2), 
\end{equation*}
for a.e.~$ \lambda \in \Lambda$ and for all ${\bf f} \in S_d(\R)$, $z\in \C$.
\end{enumerate}
Then, in the sense of Bochner integration in $H_{-q} \subset (S)'$ for a suitable $q\in \N_0$, the integral of the family of Hida distributions is itself a Hida distribution, i.e.~$\!\displaystyle \int_{\Lambda} \Phi(\lambda) \, d\nu(\lambda) \in (S)'$ and the $T$--transform interchanges with integration, i.e.~
\begin{equation*}
   T\left( \int_{\Lambda} \Phi(\lambda)  d\nu(\lambda) \right)(\mathbf{f}) =
   	\int_{\Lambda} T(\Phi(\lambda))(\mathbf{f}) \, d\nu(\lambda), \quad \mathbf{f} \in S_d(\R).
\end{equation*}
\end{corollary}

Based on the above theorem, we introduce the following Hida distribution.
\begin{definition}
\label{D:Donsker} 
We define Donsker's delta at $x \in \R$ corresponding to $0 \neq {\boldsymbol\eta} \in L_{d}^2(\R)$ by
\begin{equation*}
   \delta_0(\langle {\boldsymbol\eta},\cdot \rangle-x) := 
   	\frac{1}{2\pi} \int_{\R} \exp(i \lambda (\langle {\boldsymbol\eta},\cdot \rangle -x)) \, d \lambda
\end{equation*}
in the sense of Bochner integration, see e.g.~\cite{HKPS93,LLSW94,W95}. Its $T$--transform in ${\bf f} \in S_d(\R)$ is given by
\begin{equation*}
   T(\delta_0(\langle  {\boldsymbol\eta},\cdot \rangle-x)({\bf f}) 
   	= \frac{1}{\sqrt{2\pi \langle {\boldsymbol\eta}, {\boldsymbol\eta}\rangle}} \exp\left( -\frac{1}{2\langle {\boldsymbol\eta},{\boldsymbol\eta} \rangle}(i\langle {\boldsymbol\eta},{\bf f} \rangle - x)^2 -\frac{1}{2}\langle {\bf f},{\bf f}\rangle \right), \, \, \mathbf{f} \in S_d(\R).
\end{equation*}
\end{definition}

\subsection{Generalized Gauss Kernels}
Here we review a special class of Hida distributions which are defined by their $T$-transform, see e.g.~\cite{HS83},\cite{HKPS93},\cite{GS98a}. Proofs and more details for can be found in \cite{BG11}. Let $\mathcal{B}$ be the set of all continuous bilinear mappings $B:S_{d}(\R) \times S_{d}(\R) \to \C$. Then the functions
\begin{equation*}
S_d(\R)\ni \mathbf{f} \mapsto \exp\left(-\frac{1}{2} B({\bf f},{\bf f})\right) \in \C
\end{equation*}
for all $B\in \mathcal{B}$ are U-functionals. Therefore, by using the characterization of Hida distributions in Theorem \ref{charthm},
the inverse T-transform of these functions 
\begin{equation*}
\Phi_B:=T^{-1} \exp\left(-\frac{1}{2} B\right)
\end{equation*}
are elements of $(S)'$.

\begin{definition}\label{GGK}
The set of {\bf{generalized Gauss kernels}} is defined by
\begin{equation*}
GGK:= \{ \Phi_B,\; B\in \mathcal{B} \}.
\end{equation*}
\end{definition}

\begin{example}{\cite{GS98a}} \label{Grotex} We consider a symmetric trace class operator $\mathbf{K}$ on $L^2_{d}(\R)$ such that $-\frac{1}{2}<\mathbf{K}\leq 0$, then
\begin{align*}
\int_{S'_{d}(\R)} \exp\left(- \langle \omega,\mathbf{K} \omega\rangle \right) \, d\mu(\boldsymbol{\omega}) 
= \left( \det(\mathbf{Id +2K})\right)^{-\frac{1}{2}} < \infty.
\end{align*}
For the definition of $\langle \cdot,\mathbf{K} \cdot \rangle$ see the remark below.
Here $\mathbf{Id}$ denotes the identity operator on the Hilbert space $L^2_{d}(\R)$, and $\det(\mathbf{A})$ of a symmetric trace class operator $\mathbf{A}$ on $L^2_{d}(\R)$ denotes the infinite product of its eigenvalues, if it exists. In the present situation we have $\det(\mathbf{Id +2K})\neq 0$.
There\-fore we obtain that the exponential $g= \exp(-\frac{1}{2} \langle \cdot,\mathbf{K} \cdot \rangle)$ is square-integrable and its T-transform is given by 
\begin{equation*}
Tg({\bf f}) = \left( \det(\mathbf{Id+K}) \right)^{-\frac{1}{2}} \exp\left(-\frac{1}{2} ({\bf f}, \mathbf{(Id+K)^{-1}} {\bf f})\right), \quad {\bf f} \in S_{d}(\R).
\end{equation*}
Therefore $\left( \det(\mathbf{Id+K}) \right)^{\frac{1}{2}}g$ is a generalized Gauss kernel.
\end{example}

\begin{remark}
\begin{itemize}
\item[i)]\label{traceL2} Since a trace class operator is compact, see e.g.~\cite{RS75a}, we have that $\mathbf{K}$ in the above example is diagonalizable, i.e.~
\begin{equation*}
\mathbf{K}\mathbf{f} = \sum_{k=1}^{\infty} k_n (\mathbf{f},\mathbf{e}_n)\mathbf{e}_n, \quad \mathbf{f} \in L_d^2(\R),
\end{equation*}
where $(\mathbf{e}_n)_{n\in \N}$ denotes an eigenbasis of the corresponding eigenvalues $(k_n)_{n\in \N}$ with $k_n \in (-\frac{1}{2}, 0 ]$, for all $n \in \N$. Since $K$ is compact, we have that $\lim\limits_{n\to \infty} k_n =0$ and since $\mathbf{K}$ is trace class we also have $\sum_{n=1}^{\infty} (\mathbf{e}_n, -\mathbf{K} \mathbf{e}_n)< \infty$. We define for ${\boldsymbol \omega }\in S_d'(\R)$
\begin{eqnarray*} 
- \langle {\boldsymbol \omega }, \mathbf{K} {\boldsymbol \omega } \rangle := \lim_{N \to \infty} \sum_{n=1}^N \langle \mathbf{e}_n, {\boldsymbol \omega }\rangle (-k_n)\langle \mathbf{e}_n,{\boldsymbol \omega } \rangle. 
\end{eqnarray*}
Then as a limit of measurable functions ${\boldsymbol \omega } \mapsto -\langle {\boldsymbol \omega }, \mathbf{K} {\boldsymbol \omega } \rangle$  is measurable and hence 
\begin{eqnarray*} 
\int\limits_{S_d'(\R)} \exp(-  \langle {\boldsymbol \omega }, \mathbf{K} {\boldsymbol \omega }\rangle ) \, d\mu({\boldsymbol \omega }) \in [0, \infty].
\end{eqnarray*}
The explicit formula for the $T$-transform and expectation then follow by a straightforward calculation with help of the above limit procedure. 
\item[ii)] In the following, if we apply operators or bilinear forms defined on $L^2_d(\R)$ to generalized functions from $S'_d(\R)$, we are always having in mind the interpretation as in \ref{traceL2}.
\end{itemize}
\end{remark}
\begin{definition}\label{D:Nexp}\cite{BG11}$\;$
Let $\mathbf{K}: L^2_d(\R)_{\C} \to L^2_d(\R)_{\C}$ be linear and continuous such that:
\begin{itemize}
\item[(i)] $\mathbf{Id+K}$ is injective. 
\item[(ii)] There exists $p \in \N_0$ such that $(\mathbf{Id+K})(L^2_{d}(\R)_{\C}) \subset H_{p,\C}$ is dense.
\item[(iii)] There exist $q \in\N_0$ such that $\mathbf{(Id+K)^{-1}} :H_{p,\C} \to H_{-q,\C}$ is continuous with $p$ as in (ii).
\end{itemize}
Then we define the normalized exponential
\begin{equation}\label{Nexp}
{\rm{Nexp}}(- \frac{1}{2} \langle \cdot ,\mathbf{K} \cdot \rangle)
\end{equation}
by
\begin{equation*}
T({\rm{Nexp}}(- \frac{1}{2} \langle \cdot ,\mathbf{K} \cdot \rangle))({\bf f}) := \exp(-\frac{1}{2} \langle {\bf f}, \mathbf{(Id+K)^{-1}} {\bf f} \rangle),\quad {\bf f} \in S_d(\R).
\end{equation*}
\end{definition}

\begin{remark}
The "normalization" of the exponential in the above definition can be regarded as a division of a divergent factor. In an informal way one can write
\begin{multline*}
T({\rm{Nexp}}(- \frac{1}{2} \langle \cdot ,\mathbf{K} \cdot \rangle))({\mathbf f})=\frac{T(\exp(- \frac{1}{2} \langle \cdot ,\mathbf{K} \cdot \rangle))(\mathbf{f})}{T(\exp(- \frac{1}{2} \langle \cdot ,\mathbf{K} \cdot \rangle))(0)}\\
=\frac{T(\exp(- \frac{1}{2} \langle \cdot ,\mathbf{K} \cdot \rangle))(\mathbf{f})}{\sqrt{\det(\mathbf{Id+K})}} , \quad {\bf f} \in S_d(\R), 
\end{multline*}
i.e.~ if the determinant in the Example \ref{Grotex} above is not defined, we can still define the normalized exponential by the T-transform without the diverging prefactor. The assumptions in the above definition then guarantee the existence of the generalized Gauss kernel in \eqref{Nexp}.
\end{remark}

\begin{example}\label{pointprod}
	For sufficiently "nice" operators $\mathbf{K}$ and $\mathbf{L}$ on $L^2_{d}(\R)_{\C}$ we can define the product 
			\begin{equation*}
				{\rm{Nexp}}\big( - \frac{1}{2} \langle \cdot,\mathbf{K} \cdot \rangle  \big) \cdot \exp\big(-\frac{1}{2} \langle \cdot,\mathbf{L}\cdot \rangle \big)
			\end{equation*}
	of two square-integrable functions. Its $T$-transform is then given by 
			\begin{multline*}
				T\Big({\rm{Nexp}}( - \frac{1}{2} \langle \cdot,\mathbf{K} \cdot\rangle ) \cdot \exp( - \frac{1}{2} \langle \cdot,\mathbf{L} \cdot\rangle )\Big)({\bf f})\\
				=\sqrt{\frac{1}{\det(\mathbf{Id+L(Id+K)^{-1}})}}
				\exp(-\frac{1}{2} \langle {\bf f}, \mathbf{(Id+K+L)^{-1}} {\bf f} \rangle ),\quad {\bf f} \in S_{d}(\R),
			\end{multline*}	
	in the case the right hand side indeed is a U-functional.		
\end{example}

\begin{definition}\label{prodnexp}
Let $\mathbf{K}: L^2_{d}(\R)_{\C} \to L^2_{d}(\R)_{\C}$ be as in Definition \ref{D:Nexp}, i.e.~$${\rm{Nexp}}(- \frac{1}{2} \langle \cdot ,\mathbf{K} \cdot \rangle)$$ exists. Furthermore let $\mathbf{L}: L^2_d(\R)_{\C} \to L^2_d(\R)_{\C}$ be trace class. Then we define 
$$
{\rm{Nexp}}( - \frac{1}{2} \langle \cdot,\mathbf{K} \cdot\rangle ) \cdot \exp( - \frac{1}{2} \langle \cdot,\mathbf{L} \cdot\rangle )$$ via its $T$-transform, whenever 
\begin{multline*}
				T\Big({\rm{Nexp}}( - \frac{1}{2} \langle \cdot,\mathbf{K} \cdot\rangle ) \cdot \exp( - \frac{1}{2} \langle \cdot,\mathbf{L} \cdot\rangle )\Big)({\bf f})\\
				=\sqrt{\frac{1}{\det(\mathbf{Id+L(Id+K)^{-1}})}}
				\exp(-\frac{1}{2} \langle {\bf f}, \mathbf{(Id+K+L)^{-1}} {\bf f} \rangle ),\quad {\bf f} \in S_{d}(\R),
			\end{multline*}	
is a U-functional.
\end{definition}

In the case $\mathbf{g} \in S_d(\R)$, $c\in\C$ the product between the Hida distribution $\Phi$ and the Hida test function $\exp(i \langle \mathbf{g},. \rangle + c)$ can be defined because $(S)$ is a continuous algebra under point-wise multiplication. The next definition is an extension of this product.

\begin{definition}\label{linexp}
The point-wise product of a Hida distribution $\Phi \in (S)'$ with an exponential of a linear term, i.e.~
\begin{equation*}
\Phi \cdot \exp(i \langle {\bf g}, \cdot \rangle  +c), \quad {\bf g} \in L^2_{d}(\R)_{\C}, \, c \in \C,
\end{equation*}
is defined by 
\begin{equation*}
T(\Phi \cdot \exp(i\langle  {\bf g}, \cdot \rangle  + c))({\bf f}):= T\Phi({\bf f}+{\bf g})\exp(c),\quad {\bf f} \in S_d(\R),  
\end{equation*}
if $T\Phi$ has a continuous extension to $L^2_d(\R)_{\C}$ and the term on the right-hand side is a U-functional in ${\bf f} \in S_d(\R)$.
\end{definition}

\begin{definition}\label{donsker}
Let $D \subset \R$ with $0 \in \overline{D}$. Under the assumption that $T\Phi$ has a continuous extension to $L^2_d(\R)_{\C}$, ${\boldsymbol\eta}\in L^2_d(\R)_{\C}$, $y \in \R$, $\lambda \in \gamma_{\alpha}:=\{\exp(-i\alpha)s|\, s \in \R\}$ and that the integrand 
\begin{equation*}
\gamma_{\alpha} \ni \lambda \mapsto \exp(-i\lambda y)T\Phi({\bf f}+\lambda {\boldsymbol\eta}) \in \C
\end{equation*}
fulfills the conditions of Corollary \ref{intcor} for all $\alpha \in D$. Then one can define the product 
\begin{equation*}
\Phi \cdot \delta_0(\langle {\boldsymbol\eta}, \cdot \rangle-y),
\end{equation*}
by
\begin{equation*}
T(\Phi \cdot \delta_0(\langle {\boldsymbol\eta}, \cdot \rangle-y))({\bf f})
:= \lim_{\alpha \to 0} \int_{\gamma_{\alpha}} \exp(-i \lambda y) T\Phi({\bf f}+\lambda {\boldsymbol\eta}) \, d \lambda.
\end{equation*}
Of course under the assumption that the right-hand side converges in the sense of Corollary \ref{seqcor}, see e.g.~\cite{GS98a}.
\end{definition}

This definition is motivated by the definition of Donsker's delta, see Definition \ref{D:Donsker}.

\begin{lemma}{\cite{BG11}}\label{thelemma}
Let  $\mathbf{L}$ be a $d\times d$ block operator matrix on $L^2_{d}(\R)_{\C}$ acting component-wise such that all entries are bounded operators on $L^2(\R)_{\C}$.
Let $\mathbf{K}$ be a d $\times d$ block operator matrix on $L^2_{d}(\R)_{\C}$, such that $\mathbf{Id+K}$ and $\mathbf{N}=\mathbf{Id}+\mathbf{K}+\mathbf{L}$ are bounded with bounded inverse. Furthermore assume that $\det(\mathbf{Id}+\mathbf{L}(\mathbf{Id}+\mathbf{K})^{-1})$ exists and is different from zero (this is e.g.~the case if $\mathbf{L}$ is trace class and -1 in the resolvent set of $\mathbf{L}(\mathbf{Id}+\mathbf{K})^{-1}$).
Let $M_{\mathbf{N}^{-1}}$ be the matrix given by an orthogonal system $({\boldsymbol\eta}_k)_{k=1,\dots J}$ of non--zero functions from $L^2_d(\R)$, $J\in \N$, under the bilinear form $\left( \cdot ,\mathbf{N}^{-1} \cdot \right)$, i.e.~ $(M_{\mathbf{N}^{-1}})_{i,j} = \left( {\boldsymbol\eta}_i ,\mathbf{N}^{-1} {\boldsymbol\eta}_j \right)$, $1\leq i,j \leq J$.
Under the assumption that either 
\begin{eqnarray*}
\Re(M_{\mathbf{N}^{-1}}) >0 \quad \text{ or }\quad \Re(M_{\mathbf{N}^{-1}})=0 \,\text{ and } \,\Im(M_{\mathbf{N}^{-1}}) \neq 0,
\end{eqnarray*} 
where $M_{\mathbf{N}^{-1}}=\Re(M_{\mathbf{N}^{-1}}) + i \Im(M_{\mathbf{N}^{-1}})$ with real matrices $\Re(M_{\mathbf{N}^{-1}})$ and $\Im(M_{\mathbf{N}^{-1}})$, \\
then
\begin{equation*}
\Phi_{\mathbf{K},\mathbf{L}}:={\rm Nexp}\big(-\frac{1}{2} \langle \cdot, \mathbf{K} \cdot \rangle \big) \cdot \exp\big(-\frac{1}{2} \langle \cdot, \mathbf{L} \cdot \rangle \big) \cdot \exp(i \langle \cdot, {\bf g} \rangle)
\cdot \prod_{i=1}^J \delta_0 (\langle \cdot, {\boldsymbol\eta}_k \rangle-y_k),
\end{equation*}
for ${\bf g} \in L^2_{d}(\R,\C),\, t>0,\, y_k \in \R,\, k =1\dots,J$, exists as a Hida distribution. \\
Moreover for ${\bf f} \in S_d(\R)$
\begin{multline}\label{magicformula}
T\Phi_{\mathbf{K},\mathbf{L}}({\bf f})=\frac{1}{\sqrt{(2\pi)^J  \det((M_{\mathbf{N}^{-1}}))}}
\sqrt{\frac{1}{\det(\mathbf{Id}+\mathbf{L}(\mathbf{Id}+\mathbf{K})^{-1})}}\\ 
\times \exp\bigg(-\frac{1}{2} \big(({\bf f}+{\bf g}), \mathbf{N}^{-1} ({\bf f}+{\bf g})\big) \bigg)
\exp\bigg(-\frac{1}{2} (u,(M_{\mathbf{N}^{-1}})^{-1} u)\bigg),
\end{multline}
where
\begin{equation*}
u= \left( \big(iy_1 +({\boldsymbol\eta}_1,\mathbf{N}^{-1}({\bf f}+{\bf g})) \big), \dots, \big(iy_J +({\boldsymbol\eta}_J,\mathbf{N}^{-1}({\bf f}+{\bf g})) \big) \right).
\end{equation*}
\end{lemma}

\section{Hamiltonian Path Integrals in Momentum Space}

\subsection{The free Feynman integrand in phase space}
First we consider $V=0$ (free particle). For simplicity let $\hbar=m=1$ and $t_0=0$. Furthermore we choose to have one space dimension and one dimension for the corresponding momentum variable, i.e.~the underlying space is $S_2(\R)$. Note that higher dimensions can be obtained by multiplication of the generating functionals, since the used variables are independent. 
It is well known, see e.g.~\cite{K04} that the momentum space propagator for a free particle is given in form of a Dirac Delta function. We want to show therefore at least that we can find an expression which converges to this propagator. As above we 
consider first the action 
$$S=\int_0^t -x(\tau)\dot{p}(\tau) - \frac{1}{2m}p^2(\tau) d\tau$$
then we have with \eqref{mommom} and \eqref{mompath}
\begin{multline*}
S= \int_0^t -\sqrt{\frac{\hbar}{m}}\sqrt{\hbar m}\omega_p(\tau) \omega_x(\tau) -\frac{1}{2m} (p_0 + \frac{\sqrt{\hbar m}}{t} \langle \1_{[0,\tau)}, \omega_p\rangle)^2 \ d\tau\\
= -\hbar \int_0^t \omega_p(\tau) \omega_x(\tau) d\tau-\frac{p_0^2}{2m}t +  \langle p_0 \sqrt{\frac{\hbar}{mt^2}} (s-t) \1_{[0,t)}(s),\omega_p(s) \rangle -\frac{1}{2} \int_0^t \frac{\hbar}{t^2} \langle \1_{[0,\tau)},\omega_p\rangle^2\ d\tau
\end{multline*}
Thus we write \eqref{anpsfeymom} using $m=\hbar=1$:
\begin{multline*}
{\rm Nexp}\bigg(-\frac{1}{2} \big\langle
(\omega_x,\omega_p),K_{mom}(
    \omega_x,
    \omega_p )\big\rangle\bigg)\\
    \times \exp(\langle p_0 \frac{1}{t} (\cdot-t) \1_{[0,t)}(\cdot),\omega_p\rangle) \delta(p'-p_0 - \langle \frac{1}{t} \1_{[0,t)}, \omega_p\rangle)
\end{multline*}
where the operator matrix $K$ on $L_2^2(\R)_{\C}$ can be written as
\begin{equation}\label{kinmatmom}
K_{mom}=\left(
\begin{array}[h]{l l}
    -\1_{[0,t)}&{i} \1_{[0,t)}\\[0,1 cm]
    {i}\1_{[0,t)}& - \1_{[0,t)} +\frac{i}{t^2}A
\end{array}
\right).
\end{equation}
Here $$A \,f(s)=\1_{[0,t)}(s) \int_s^t \int_0^{\tau} f(r) \, dr \, d\tau, f \in L^2(\R,\C), s\in \R,$$ we refer to \cite{GS98a} for properties of the operator as the trace class property, invertibility and spectrum.\\
We have for $f,g \in L^2(\R)_{\C}$
$$\langle f, A g\rangle = \int_0^t \int_0^{\tau} f(s) \, ds \cdot \int_0^{\tau} g(s) \, ds\ d\tau.$$
Hence the operator is used to implement the integral over the squared Brownian motion. The last term pins the momentum variable to $p'$ at $t$. Note that the space variable is not pinned.\\
Our aim is to apply Lemma \ref{thelemma} with $K$ as above and ${\bf g}=(0,\frac{p_0}{t} (s-t) \1_{[0,t)}(s))$, $L=0$ and as ${\boldsymbol\eta}= (0,\frac{1}{t}\1_{[0,t)})$.
The inverse of $(Id+K)$ is given by
\begin{eqnarray}
N^{-1}=(Id+K_{mom})^{-1}=\label{InvNmom}
\bigg(
\begin{array}{l l}
 \1_{[0,t)^c} & 0\\
    0& \1_{[0,t)^c}
\end{array} 
\bigg)+i\bigg(
\begin{array}{l l}
 \frac{1}{t^2} A & -\1_{[0,t)}\\
    -\1_{[0,t)}& 0
\end{array} 
\bigg),
\end{eqnarray}
hence $({\boldsymbol\eta},N^{-1}{\boldsymbol\eta})=0$.\\ 
To apply Lemma \ref{thelemma} we use a small perturbation of the matrix $N^{-1}$.\\
Let $\epsilon >0$ then we define 
$$N^{-1}_{\epsilon} =\bigg(
\begin{array}{l l}
 \1_{[0,t)^c} & 0\\
    0& \1_{[0,t)^c}
\end{array} 
\bigg)+\bigg(
\begin{array}{l l}
 \frac{i}{t^2} A & -i\1_{[0,t)}\\
    -i\1_{[0,t)}& +\epsilon 
\end{array} 
\bigg),$$
then we have $({\boldsymbol\eta},N_{\epsilon}^{-1}{\boldsymbol\eta})=\frac{\epsilon}{t}$ and Lemma \ref{thelemma} can be applied. 
Therefore the assumptions of Lemma \ref{thelemma} are fulfilled. 
Thus we define the regularized free momentum integrand by its $T$-transform in $(f_x,f_p) \in S_2(\R)$
\begin{multline*}
T(I_{0mom,\epsilon})(f_x,f_p)  =\frac{1}{\sqrt{2 \pi \frac{\epsilon}{t}}} \exp(-\frac{ip_0^2}{2}t)\\ \times \exp\Big(-\frac{1}{2} \left\langle \left(\begin{array}{c} f_x \\ f_p + \frac{p_0}{t} (\cdot-t) \1_{[0,t)}\end{array} \right) ,N^{-1}_{\epsilon} \left(\begin{array}{c} f_x \\ f_p +\frac{p_0}{t} (\cdot-t) \1_{[0,t)}\end{array} \right)\right\rangle\Big)\\
\times \exp\Bigg(\frac{1}{2\frac{\epsilon}{t}} \bigg(i(p'-p_0) +\Bigg\langle \left(\begin{array}{c} f_x \\ f_p +(\cdot-t) \1_{[0,t)}\end{array} \right), N^{-1}_{\epsilon} \left(\begin{array}{c} 0 \\ \1_{[0,t)}\end{array} \right)\Bigg\rangle\bigg)^2\Bigg)
\end{multline*}
Hence its generalized expectation 
\begin{multline*}
\mathbb{E}(I_{0,mom,\epsilon}) = TI_{0,mom,\epsilon}(0) \\
=\frac{\sqrt{t}}{\sqrt{2 \pi \epsilon}}  \exp(-\frac{1}{2} \left\langle \left(\begin{array}{c} 0\\ 0 + \frac{p_0}{t} (\cdot-t) \1_{[0,t)}\end{array} \right), N^{-1}_{\epsilon} \left(\begin{array}{c} 0\\ \frac{p_0}{t} (\cdot-t) \1_{[0,t)}\end{array} \right)\right\rangle)\\
\times \exp\left(\frac{t}{2{\epsilon}} \Bigg(i(p'-p_0) +\Bigg\langle \left(\begin{array}{c} 0 \\0+\frac{p_0}{t} (\cdot-t) \1_{[0,t)}\end{array} \right), N^{-1}_{\epsilon} \left(\begin{array}{c} 0 \\ \frac{1}{t}\1_{[0,t)}\end{array} \right)\Bigg\rangle\Bigg)^2 \right)\cdot \exp(-\frac{ip_0^2}{2}t)\\
=\frac{\sqrt{t}}{\sqrt{2 \pi \epsilon}}  \exp\left(-\frac{\epsilon}{2t^2}p_0^2  \int_0^t (s-t)^2 \ ds\right)\\
\times\exp\left(\frac{t}{2{\epsilon}} \big(i(p'-p_0) +\frac{p_0\epsilon}{t^2} \int_0^t (s-t)\, ds\big)^2\right) \exp(-\frac{ip_0^2}{2}t)\\
=\frac{\sqrt{t}}{\sqrt{2 \pi \epsilon}}  \exp\left(-\frac{\epsilon}{2t^2}p_0^2  \int_0^t (s-t)^2 \ ds\right) \exp(-\frac{ip_0^2}{2}t)\\
\times \exp\left(\frac{t}{2 {\epsilon}} \bigg(-(p'-p_0)^2 +\frac{2i p_0\epsilon}{t^2}(p'-p_0) (\int_0^t (s-t)\, ds)+ \frac{p_0^2\epsilon^2}{t^4} \big(\int_0^t (s-t)\, ds\bigg)^2\right) \\
=\frac{\sqrt{t}}{\sqrt{2 \pi \epsilon}}  \exp(-\frac{t}{2\epsilon} (p'-p_0)^2)\exp(-\frac{ip_0^2}{2}t)\\
\times \exp\left(-\frac{\epsilon}{2t^2}p_0^2  \int_0^t (s-t)^2 \ ds\right) \exp\left(\frac{p_0^2\epsilon}{2t^3} \big(\int_0^t (s-t)\, ds\big)^2\right) \cdot \exp(i\frac{p_0}{t} (p'-p_0)\int_0^t (s-t)\, ds)
\end{multline*}
In the limit $\epsilon \to 0$ we obtain:\begin{eqnarray*}
\lim_{\epsilon \to 0} \mathbb{E}(I_{0mom,\epsilon}) = \delta(p'-p_0)\cdot \exp(-\frac{ip_0^2}{2}t) \cdot \exp\big(\frac{ip_0}{t} (p'-p_0)\int_0^t (s-t)\, ds\big)\\
= \delta(p'-p_0)\cdot \exp(-\frac{ip_0^2}{2}t),
\end{eqnarray*}
Note that the Delta function just gives values if $p'=p_0$ and thus serves to conserve the momentum of the free particle. If there is no potential the momentum must be the same as the initial momentum since the space is free of any force. Moreover the generalized expectation yields the propagator of the free particle in momentum space, see \cite{K04}.  

\subsection{Harmonic oscillator in Momentum Space}
In this section we construct the Feynman integrand for the harmonic oscillator in phase space. I.e.~the potential is given by $x \mapsto V(x)= \frac{1}{2}k x^2$, $0 \leq k<\infty$.
The corresponding Lagrangian in phase space representation in momentum space is given by
\begin{equation*}
(x(\tau), p(\tau)) \mapsto L((x(\tau), p(\tau)))=-\dot{p}(\tau) {x}(\tau)-\frac{p(\tau)^2}{2} - \frac{1}{2}k x(\tau)^2.
\end{equation*}
In addition to the matrix $K$ from the free case, see \eqref{kinmatmom}, we have a matrix $L$ which includes the information about the potential, see also \cite{GS98a} and \cite{BG11}. For the sake of simplicity we set $p_0 =0$.
In order to realize \eqref{anpsfeymom} for the harmonic oscillator we consider
\begin{multline*}
I_{HO,mom}={\rm Nexp}\big(-\frac{1}{2} \langle (\omega_x,\omega_p), K_{mom} (\omega_x,\omega_p) \rangle \big) \cdot \exp\big(-\frac{1}{2} \langle (\omega_x,\omega_p), L (\omega_x,\omega_p) \rangle\big)\\
 \cdot \delta\big(\langle (\omega_x,\omega_p), (0,\1_{[0,t)}) \rangle -(p')\big).
\end{multline*}
with
\begin{equation*}
L=\left(
\begin{array}{l l}
i kt^2 \1_{[0,t)} & 0\\
0 & 0
\end{array}
\right),\, p' \in \R, \,t>0.
\end{equation*}

Then we have
$$N=(Id+K+L) = \left(\begin{array}{ll} \1_{[0,t]^c}& 0 \\ 0 &\1_{[0,t)^c} \end{array} \right) + i\left(\begin{array}{ll} kt^2\1_{[0,t]}& \1_{[0,t)} \\ \1_{0,t)} &\frac{1}{t^2} A \end{array} \right).$$

Its inverse is then given by
$$N^{-1}=(Id+K+L)^{-1} = \left(\begin{array}{ll} \1_{[0,t]^c}& 0 \\ 0 &\1_{[0,t)^c} \end{array} \right) + \frac{1}{i} \left(\begin{array}{ll} \frac{A}{t^2}(kA-\1_{[0,t)})^{-1}& -(kA-\1_{[0,t)})^{-1} \\-(kA-\1_{[0,t)})^{-1} &kt^2(kA-\1_{[0,t)})^{-1} \end{array} \right),$$
if $(kA- \1_{[0,t)})^{-1}$ exists, i.e.~$kA- \1_{[0,t)}$ is bijective on $L^2_2([0,t))$. 
The eigenvalues $l_n$ of $A$, which are different from zero yield: 
\begin{equation*}
l_n=k  \bigg(\frac{t}{(n-\frac{1}{2})\pi}\bigg)^2, \quad n \in \N.
\end{equation*}
Thus $(kA- \1_{[0,t)})^{-1}$ exists if $l_n \neq 1$ for all $n\in \N$. 
Which is true for $0<t< \infty$ $t\neq \frac{(2m-1)\pi}{(2\sqrt{k})}, m \in \N$. 
Hence we obtain using \cite[p.~431, form.~1]{GR65}: 
\begin{align*}
\frac{1}{\det(Id +L(Id+K)^{-1})}
&=\det\left( Id + \left(\begin{array}{l l}
-k A & -kt^2\1_{[0,t)}\\
0&0 
\end{array}
\right)\right)^{-1}\\
&= \big(\prod_{n=1}^{\infty} (1-k \big(\frac{t}{(n-\frac{1}{2})\pi}\big)^2)\big)^{-1}=\frac{1}{\cos(\sqrt{k} t)}.
\end{align*} 

We use the eigenstructure of $A$ to determine the matrix $M_{N^{-1}}$. We have
\begin{multline*}
M_{N^{-1}} = \langle (0, \frac{1}{t} \1_{[0,t)}), N^{-1} (0, \frac{1}{t} \1_{[0,t)})\rangle
=\frac{ikt^2}{t^2} \sum_{n=0}^{\infty} \frac{1}{1-l_n} \langle e_n, \1_{[0,t)}\rangle^2 \\=ik \frac{\tan(\sqrt{k}t)}{\sqrt{k}} = i\sqrt{k} \tan(\sqrt{k}t).
\end{multline*}
We see that due to our assumptions on $t$, the requirements to apply Lemma \ref{thelemma} are fulfilled, since $M_{N^{-1}}$ is completely imaginary with positive imaginary part.

The $T$-transform of $I_{HO,mom}$ in ${\bf f} \in S_2(\R)$ with ${\boldsymbol \eta}= (0,\frac{1}{t}\1_{[0,t)})$ is then given by
\begin{multline}\label{genfunImomHO}
T I_{HO}({\bf f})= \sqrt{\left(\frac{1}{2\pi i \sqrt{k} \sin(\sqrt{k} t)}\right)} \\
\times\exp\!\left(\frac{1}{2} \frac{1}{i\sqrt{k}\tan(\sqrt{k} t)} \Big(ip'+\big\langle{\boldsymbol{\eta}},\frac{1}{i} \left(\begin{array}{ll} \frac{A}{t^2}(kA-\1_{[0,t)})^{-1}& -(kA-\1_{[0,t)})^{-1} \\-(kA-\1_{[0,t)})^{-1} &kt^2(kA-\1_{[0,t)})^{-1} \end{array} \right) ({\bf f} +{\bf g})\big\rangle \Big)^2\right)\\
\times\exp\!\Bigg(-\frac{1}{2} \bigg( \big({\bf f} + {\bf g}\big) ,\! 
 \left(
\begin{array}{l l}
\1_{[0,t)^c} & 0 \\
0 & \1_{[0,t)^c} \end{array}\right)
\big({\bf f} + {\bf g}\big) \bigg)\!\Bigg)\\
\times\exp\!\Bigg(-\frac{1}{2} \bigg( \big({\bf f} + {\bf g}\big) ,\! 
 \frac{1}{i} \left(\begin{array}{ll} \frac{A}{t^2}(kA-\1_{[0,t)})^{-1}& -(kA-\1_{[0,t)})^{-1} \\-(kA-\1_{[0,t)})^{-1} &kt^2(kA-\1_{[0,t)})^{-1} \end{array} \right)
\big({\bf f} + {\bf g}\big) \bigg)\!\Bigg)
\end{multline} 

Finally we obtain the following theorem. 

\begin{theorem}\label{hothmmom}
Let $y\in \R$,  $0<t< \infty$ $t\neq \frac{(2m-1)\pi}{(2\sqrt{k})}, m \in \N$, then the Feynman integrand for the harmonic oscillator in phase space in momentum space $I_{H0,mom}$ exists as a Hida distribution and its generating functional is given by \eqref{genfunImomHO}. Moreover its generalized expectation
\begin{equation*}
\mathbb{E}(I_{HO,mom})=T(I_{HO,mom})(0)=\sqrt{\left(\frac{1}{2\pi i \sqrt{k} \sin(\sqrt{k} t)}\right)} \exp\left( i \frac{1}{2\sqrt{k}\tan(\sqrt{k} t)} p'^2\right)
\end{equation*}
is the Greens function to the Schrö\-dinger equation for the harmo\-nic oscil\-lator in momentum space, compare e.g.~with \cite[p.118, form.2.187]{K04}.   
\end{theorem}


%

\end{document}